# Predicted annual energy yield of III-V/c-Si tandem solar cells: modelling the effect of changing spectrum on current-matching

Ian Mathews, Shenghui Lei, Ronan Frizzell

*Bell Labs, Nokia, Dublin, Ireland*

**Abstract**

High efficiencies of >30% are predicted for series-connected tandem solar cells when current-matching is achieved between the wide-bandgap top cell and silicon bottom cell. Sub-cells are typically optimised for current-matching based on the standard AM1.5G spectrum, but in practice, the incident radiation on a solar cell can be very different from this standard due to the effects of the sun's location in the sky, atmospheric conditions, total diffuse element etc. The resulting deviations in spectral content from optimum conditions lead to current mismatch between tandem cell layers that adversely affects the device's performance. To investigate the impact of this issue the energy yield (%) of tandem solar cells comprising a III-V wide-bandgap solar cell connected electrically and optically in series with a silicon bottom cell was simulated over a full year using measured spectral data from Denver, CO. Top cells with bandgaps from 1.5-1.9 eV were modelled using an external radiative efficiency method. The predicted annual energy yields were as high as 31% with an optimum 1.8 eV top cell, only 2.8% lower (absolute) than the AM1.5G predicted efficiency. The annual energy yield of tandem cells with no current-matching constraint, i.e. parallel-connected devices, was also simulated. Here the difference between series and parallel connections were only significant for non-optimum bandgap combinations. Our results indicate that AM1.5G based optimization of sub-cells can be effectively employed to achieve high energy yields of >25% for III-V/Si tandem solar cells in mid-latitude US locations, despite the continuous variation in spectra throughout a calendar year.

## 1. Introduction

Silicon solar cells first demonstrated by Bell Laboratories in the 1960's have become the dominant PV technology of the last number of decades. While progress has been made in improving the efficiency and reducing the cost of thin-film technologies such as CIGS, CdTe and Perovskites [1], single-junction silicon solar cells accounted for 90% of globally installed PV in 2016 [2]. Industrial research labs have recently demonstrated silicon solar cells, produced using pilot manufacturing lines, with efficiencies of greater than 25% [3], which represents the near-future capabilities of this technology. The theoretical maximum efficiency for single-junction silicon solar cells is 29.4% while a more practical limit is argued to be 27%, with the primary causes of the difference being emitter recombination, rear-surface optical loss and edge losses [4]. Therefore single-junction silicon cells are reaching their performance peak with a clear strategy to put high-efficiency technologies into production over the coming years [5]. Efficiency is not the only performance metric for solar cells with reduced costs key to achieving low $/W modules with reduced material usage being achieved by using thinner silicon wafers. Nevertheless, a number of new cell architectures are under investigation to boost silicon solar cells performance above 30% and significantly reduce $/W. The majority of these look to combine an additional pn-junction(s) with the primary silicon cell to form tandem cell architectures, mimicking the high-performance III-V multi-junction solar cells used in concentrated photovoltaic systems (CPV) [6,7]. Strategies to form these cells typically provide a wide-bandgap top cell, on the front surface of the silicon sub-cell, in the form of a-Si [8], Perovskite [9], III-V solar cells [10] or nanowires [11]. Significant efficiencies of close to 30% under 1-sun illumination have now been achieved with dual-junction cells that combine a III-V top cell and silicon bottom cell [10].

Silicon-based tandem cells incorporating single-junction III-V or Perovskite top cells are a strong possibility to advance commercial solar cells towards efficiencies of 30%. A primary consideration to optimise the power produced by a tandem solar cell is the bandgap combination of the materials used [11]. The Shockley-Quiesser detailed balance method has been used to show the optimum silicon-based tandem cell bandgap combination under AM1.5G is 1.72 eV and 1.12 eV [12] with predicted idealised efficiencies of up to 45%. It has been shown that considering the real electrical/optical properties of silicon solar cells, versus an idealised 1.1 eV absorber, the optimum bandgap is 1.75 – 1.82 eV and varies as a function of the silicon cell quality where narrower bandgap top cells can be used with higher performing silicon cells [13]. These high efficiencies predicted for series-connected tandem cells are expected due to the optimum current matching achieved between the two sub-cells under a single spectrum. In practice the incident radiation on a solar cell varies from minute to minute due to the suns location in the sky, atmospheric conditions, total diffuse element, etc. For series-connected cells this is a problem as any alteration from AM1.5G leads to an imbalance in the currents produced by each cell and loss of current-matching. As the total photo current produced by the tandem solar cell is the lesser of any junction (Kirchoff's law), in practice the efficiency of a tandem cell may vary significantly due to changes in the incident spectrum. To check the suitability of AM1.5G based optimisation of tandem cells, we model the efficiency of silicon-based tandem solar cells comprising a III-V top cell using this standard spectrum and compare it to the predicted annual energy yield considering real spectral data measured in Denver, Colorado. It is noted that AM1.5G is a representative spectrum for certain times of the year at mid-latitudes in the USA (incl. Denver) due to the input conditions used to derive the AM1.5G standard. However, the actual incident spectrum in Denver will deviate considerably from the AM1.5G standard throughout the year and they will rarely, if ever, be equal. These continuous deviations in spectral content will lead to current mismatch between tandem cell layers and this paper will quantify the effects of this over the course of a calendar year.

The annual energy yield (%) of single-junction cells has previously been examined using representative hour-by-hour spectra derived from historical meteorological data and it has been shown it is typically 1-2% (absolute) lower than the AM1.5G predicted efficiency [14]. Triple-junction solar cells for CPV systems, with layers optimised for AM1.5d conditions, have previously been shown to suffer a ~10% relative difference between predicted efficiency and predicted energy yield due to the effects of variation in spectral content at different deployment locations [15]. We instead focus on providing understanding of the response of dual-junction devices to measured spectral conditions, similar to previous work on the GaAs-Si [14] material system but expanded to consider the effect of bandgap of the top cell. We compare a number of top cell bandgaps over a 1.5 – 1.9 eV range and show that over a full year the optimum top-cell bandgap remains ~1.8 eV based on our predictions of the annual energy yield from the cells. The percentage energy yield over the year is less than the efficiency predicted under the AM1.5G spectrum with the yield of each bandgap combination 2.4-2.8% (absolute) less than the AM1.5G predicted efficiency. It is also noted that the differences in the annual energy yield under measured spectra and the predicted efficiency under the AM1.5G standard spectrum does not vary significantly as a function of top cell bandgap. Finally we show that for an optimum top cell bandgap, the difference between the energy produced over the year by providing parallel versus series connections to the sub-cells is minimal, but for non-optimum bandgap combinations extra terminals significantly increase the yield. The results are specific to the year the measured spectrum was recorded and deviations in measured spectrum would be expected year on year. However, the results provide important insight into how the annual energy yield can vary under real illumination conditions and highlights the effects this can have on tandem cells optimised using the AM1.5G standard spectrum.

## 2. Method

### 2.1. *Tandem cell model*

The solar cell simulated was a dual-junction device consisting of a III-V top cell with a bandgap ($E_g$) varying from 1.5 – 1.9 eV connected electrically and optically in series with a silicon bottom cell by an idealised, fully transparent, low resistance tunnel junction (Figure 1). The overall goal of the model was to determine the current-voltage (*J-V*) behaviour of the tandem cell for any given spectrum of solar irradiance. As Kirchoff's law limits the photocurrent produced by two cells connected in series to the smaller produced by each, it was necessary for the model to determine the minimum photocurrent of the two cells and to use this to calculate the overall *J-V* behaviour.

While detailed balance theory provides a useful tool for general bandgap optimisation, it assumes 100% absorption of photons with energy greater than the semiconductor bandgap in each cell. Given the indirect optical absorption of silicon, it would over-estimate the current generation in the silicon bottom-cell in this case, leading to incorrect estimation of the current-match/mismatch conditions in the tandem cell. Instead, both layers of the solar cell were modelled using the one-dimensional optical and electrical properties of the materials considered.

The *J-V* behaviour of each cell is approximated using the standard two-diode model (Equation 1). The total current in each cell is the light-generated current density, $J_L$, less the recombination currents, $J_{o1}$ & $J_{o2}$, which are the radiative recombination and non-radiative recombination in the bulk and depletion regions respectively. Typically $J_{o1}$ dominates at higher biases (close to the operating voltage) with an ideality close to one, while $J_{o2}$ dominates at lower biases with an ideality close to two; this variation in ideality is accounted for in the application of Equation 1.

$$J_{Total}(V) = J_L - J_{o1}(e^{qV/kT} - 1) - J_{o2}(e^{qV/2kT} - 1) \qquad (1)$$

The light-generated current density, $J_L$, in each cell was found using Equation (2). The photon flux (*PF*) incident on the device was first determined according to $PF = P_o\lambda/hc$, where $P_o$ is the initial intensity of light at wavelength $\lambda$, $h$ is Plank's constant and $c$ is the speed of light. The absorbed photons were then calculated using the Beer-Lambert method, considering single pass absorption (no back mirror) with optical losses due to front-grid metal shading loss, front-surface reflectance and reflectance between the cells was neglected for all wavelengths. The absorption coefficient, $\alpha$, required for Equation (2), was modelled separately for each cell of the tandem device; $\alpha$ for the *top cell* was modelled as a function of $E_g$ using Equation (3), which was experimentally derived by Kurtz *et al.* for $Ga_xIn_{1-x}P$ [16], while $\alpha$ for the silicon *bottom cell* was found using the data provided by Green *et al.* [17]. For the bottom cell, $J_L$ was found considering the spectrum transmitted through the top cell, i.e. photons absorbed in the III-V top cell did not affect the bottom cell. The cell thickness was accounted for using *x* in Equation 2.

In this study we are interested in the effect of real spectral conditions on the annual energy yield of multi-junction solar cells for 1-sun applications and opted not to include temperature effects in our analysis. This will result in an overestimation of the cell efficiency at some points throughout the year, as the higher operating cell temperatures during periods of more intense radiation would be expected to only marginally reduce the efficiency of the tandem cell. However, this trade-off was considered acceptable for this work since the focus here is on an analysis of the current matching conditions in the tandem cells with varying bandgap III-V top cells. While consideration of the effect of temperature on optical absorption could have aided the study, this was prevented by a lack of reliable temperature dependent data for the full range of III-V cells considered in this work. It is expected that elevated operating temperatures will increase the absorption edge of the Si sub-cell more readily than the III-V cells, which could adversely affect the current matching. However, the temperature changes involved in 1-sun applications are not expected to be sufficient to cause significant deviations from the current-matching conditions described later in the paper. Based on this analysis, the tandem cell temperature was fixed at 298K throughout this study.

$$J_L = \int_{\lambda=346\,nm}^{\lambda=1066\,nm} q.PF(\lambda)(1 - e^{-\alpha(\lambda)x}) \qquad (2)$$

$$\alpha(E) = 4.55(E - E_g)^{1/2} + 2.05(E - E_g - 0.1)^{1/2} \quad [298\ K] \quad (3)$$

For the III-V top cells we assume high-quality single-crystal devices where radiative recombination will dominate. The recombination currents were modelled using the radiative efficiency method developed by Chan *et al.* and described in [18] and summarised here. The radiative recombination rate, $J_{o1}$, can be found for a general semiconductor using a generalised form of Planck's equation and applying the Boltzmann approximation with the resulting photon flux converted to current. This is described by Equation (4), where *n* is the refractive index that was set to 3.6 for all top cell materials and *k* is Boltzmann's constant.

$$J_{o1} = \frac{2q\pi n^2 kT}{4\pi h^3 c^2} e^{-E_g/kT}(E_g^2 + 2E_g kT + 2k^2 T^2) \quad (4)$$

The radiative efficiency of the III-V top cell (Equation 5) was introduced to determine the value of $J_{o2}$ in Equation 1. By setting the radiative efficiency to 10%, our model matches what has been achieved in state-of-the-art III-V/Si tandem solar cells using wafer bonding [19]. Integrating III-V solar cells on silicon using metamorphic growth, or alternative integration methods will likely result in lower radiative efficiencies, however, we use 10% throughout this study to consider the current state-of-the-art. This whole approach allows us to compare III-V cells of different bandgaps without empirically derived current transport properties. The value of $J_{o2}$ evaluated using Equation (5) can be used in Equation (1) to determine the *J-V* behaviour of the top cell.

$$\eta_{rad}(V) = \frac{J_{o1}(e^{qV/kT} - 1)}{J_{o1}(e^{qV/kT} - 1) + J_{o2}(e^{qV/2kT} - 1)} \quad (5)$$

Silicon is an indirect bandgap material with low radiative recombination rates in lowly doped samples. Therefore, the dominant recombination mechanisms in silicon solar cells are non-radiative and for this reason we did not use the radiative efficiency method described above (in Equation 4) to model the recombination currents in the silicon bottom cell. Instead, the electrical properties of the silicon cell were modelled as a state of the art c-Si PERL cell with a 1-sun AM1.5G efficiency of 25% [20]. The current-voltage characteristics are simulated using a single-diode model with a recombination current, $J_o$, of 4.9 x $10^{-11}$ A/cm$^2$ and an ideality of 1, which gives the reported performance under standard illumination [21]. In our model the short-circuit current is found using Equation (2) considering the spectrum transmitted by the top cell, the absorption coefficient of Si, and considering single-pass absorption, as the simplest III-V integration strategies will likely require a flat Si surface, and a cell thickness of 150 μm, that is thinner than the 400 μm required to achieve 25% efficiency in the state-of-the-art PERL cell, but represents a value the industry is moving towards to reduce material costs and is likely to be common by the time tandem cells are commercialised [2]. The simulated efficiency of this Si solar cell under 1-sun AM1.5G, not in tandem configuration, was 21.7% under 1000 W/m$^2$ AM1.5G with a short-circuit current density of 38.9 mA/cm$^2$ and an open-circuit voltage of 657 mV.

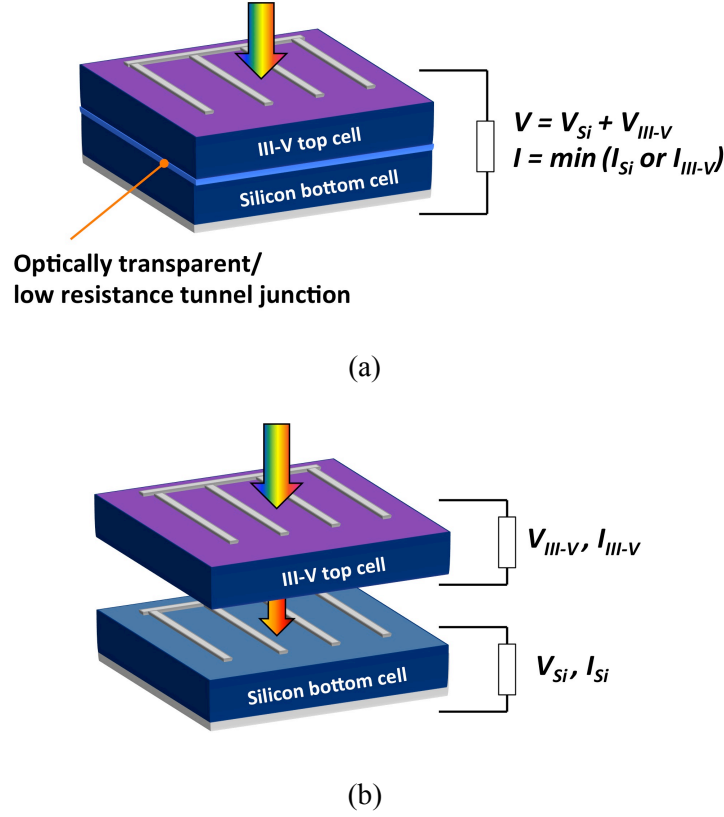

Figure 1: Schematic cross-sections of the (a) 2-terminal series-connected and (b) 4-terminal parallel connected tandem cell considered.

### 2.2. *Modelling Annual energy yield*

The energy yield of a tandem cell structure over a calendar year was predicted by simulating the efficiency of the device under spectral data measured at the NREL's Solar Radiation Research Laboratory Baseline Measurement System in Denver, Colorado [22]. The lab is located at Latitude 39.74º North, Longitude 105.18º West at an elevation of 1829 metres. The data was measured using an EKO MS-700 Spectroradiometer with traceability to NIST Standard of Spectral Irradiance Lamp Standards. The spectral range of the measurements is 346 to 1066 nm taken at 3.3 nm intervals every 60 seconds. Measurements are taken at a 40º south tilt on the roof i.e. a flat-plate (no-tracking), south-facing and latitude tilt-orientation. The data used in this study was measured from the 1st January 2014 to 31st December 2014.

The model sequence ran as follows: for each minute of the year, the photocurrent produced by the top cell, $J_{L-top}$, was found considering its wavelength dependent absorption coefficient (Equation 1) and the incident spectrum at that time. Subsequently the photocurrent produced by the silicon cell, $J_{L-Si}$, was found considering the spectrum transmitted after absorption in the top cell and the wavelength dependent absorption coefficient of silicon, as explained in the previous section. Five top cell bandgaps of 1.5, 1.6, 1.7, 1.8 and 1.9 eV were considered. As Kirchoff's law limits the photocurrent produced by two solar cells connected in series to the smaller produced by each, $J_L$ was taken as the minimum of $J_{L-top}$ and $J_{L-Si}$. A *J-V* curve was derived for the top cell according to Equation 3, using $J_{o1}$ and $J_{o2}$ (from Equations 4 & 5) and combined with the *J-V* curve of the silicon cell. These *J-V* curves were added in series and the maximum power point of the resulting curve was calculated to determine the instantaneous power produced in W/m². This was completed for the full year using a time-step of 1 minute with the total energy produced found by integrating the power over the year according to Equation (6).

$$Energy\ production\ (kWh/m^2) = \int_{t_1}^{t_2} (J_{MPP}\, V_{MPP})\, dt \qquad (6)$$

The monthly and annual energy yields (%) were found, according to Equation (7), by dividing the energy produced by each tandem design by the total measured solar energy incident on a 40º surface given in [22] rather than integrating the spectra used. This was because all the spectra used for simulations terminated at 1100 nm and do not include IR radiation, therefore integrating these spectra would underestimate the incident solar energy and overestimate the yield.

$$Energy\ yield\ (\%) = 100 \times \frac{Energy\ produced\ (kWh/m^2)}{Energy\ available\ (kWh/m^2)} \qquad (7)$$

3. Results

    *3.1. Efficiency under AM1.5G*

Before calculating the annual energy yield from the various bandgap combinations, the top cell thickness was optimised to current match with a silicon bottom cell, with the optimisation carried out considering the AM1.5G spectrum. For the wider bandgaps considered (>1.8 eV) it was found > 3 µm of material was required to achieve current-matching. In these cases, we limited the top cell thickness to 3 µm as we consider thicker materials will be too costly to be practical. In addition, the diffusion lengths in wide-bandgap III-V materials would not be sufficient to allow the cell to be thicker than 3 µm without significant impact on the cell Voc. Although the 3 µm thickness would have negative effects on the top cell Voc, the extent of the impact would vary for different materials. Including specific descriptions of the carrier transport behaviour of each III-V top cell considered would limit the generality of the model and prevented direct comparison of the different bandgap top cells, which was the main intention of the paper. The thicknesses of the top cell layer and predicted tandem cell efficiency under AM1.5G for a range of bandgap combinations are given in Figure 2. For the cases where the thickness was limited to 3 µm, the top cell cannot generate the necessary current to balance the Si bottom cell and so the overall tandem cell performance is current-limited by the top-cell. The photovoltaic properties of the 5 devices analysed in the remainder of this paper are highlighted in Table I.

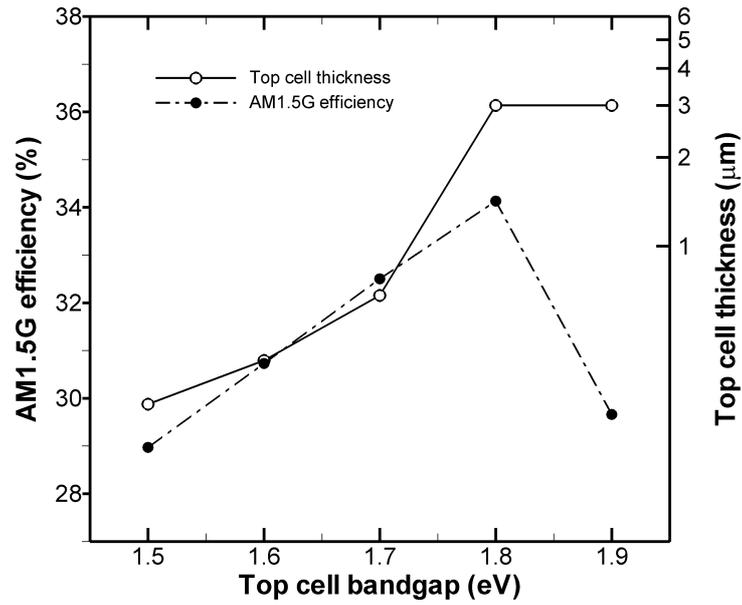

Figure 2: Predicted efficiency of a series-connected III-V/Si tandem solar cell as function of top cell bandgap. For each bandgap considered the thicknesses used for the top cell is shown. The thickness of the Si bottom cell was 150 μm.

Table I: Simulated photovoltaic properties under AM1.5G of the tandem solar cells considered.

| $E_g$ (eV) | $V_{oc}$ (V) | $J_L$ (mA/cm$^2$) | FF (%) | η (%) |
|---|---|---|---|---|
| 1.5 | 1.79 | 19.46 | 82.50 | 28.97 |
| 1.6 | 1.89 | 19.46 | 82.93 | 30.73 |
| 1.7 | 1.98 | 19.46 | 83.72 | 32.50 |
| 1.8 | 2.08 | 19.26 | 84.60 | 34.13 |
| 1.9 | 2.04 | 16.60 | 86.81 | 29.66 |

3.2. *Annual energy yield*

The predicted annual energy yields (AEY) over the entire year for each tandem cell under the measured irradiation are given in Table II and compared to the AM1.5G efficiency predicted in the previous section. The energy available to a 40º tilt solar module at this location in 2014 was 1932 kWh/m$^2$. Table II shows that the predicted annual energy yield and AM1.5G efficiency differ by only 2.4 - 2.8% (absolute) over the top cell bandgap range. The energy yield versus top cell bandgap follow the trend of the AM1.5G efficiency; 1.8 eV remains the optimum top cell, while all other bandgaps considered maintain the same ranking in terms of energy produced whether irradiated by AM1.5G or the measured spectra.

Table II: Annual energy yield (kWh/m² and %) of the five tandem solar cells considered in the simulations.

| Top cell $E_g$ | (eV) | 1.5 | 1.6 | 1.7 | 1.8 | 1.9 |
|---|---|---|---|---|---|---|
| Energy produced | (kWh/m²) | 514 | 544 | 575 | 605 | 523 |
| Annual energy yield* | (%) | 26.58 | 28.18 | 29.78 | 31.3 | 27.07 |
| AM1.5G efficiency | (%) | 28.97 | 30.73 | 32.50 | 34.13 | 29.66 |
| AEY – AM1.5G | (%) | 2.39 | 2.55 | 2.72 | 2.83 | 2.59 |

* Energy available = 1932 kWh/m²

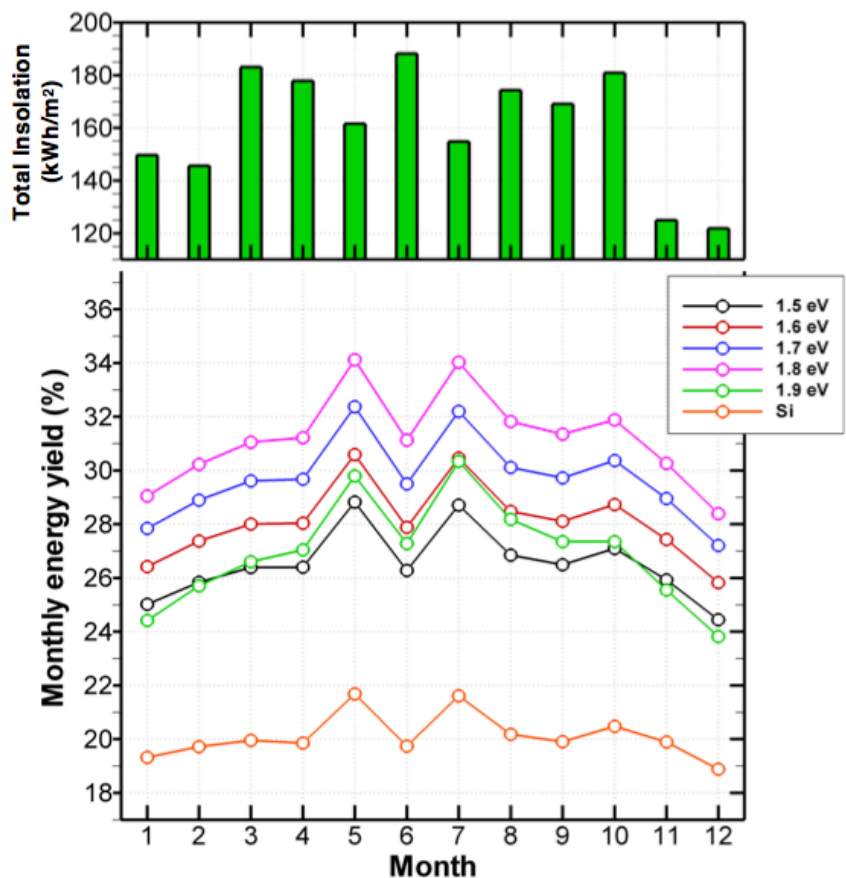

Figure 3: The predicted monthly energy yield (%) of the tandem solar cells with the results for a single-junction silicon solar cell given for comparison. The monthly total insolation is given (top axis) to show the energy yield is not directly proportional to the total irradiance.

The monthly energy yield (%) for each tandem cell is given in Figure 3 and shows that 1.8 eV remains the optimum top cell bandgap each month, despite variations in the monthly spectra. From 1.5 eV to 1.8 eV, the energy yield during each month increases with increasing bandgap. Also shown at the top of the figure is the total irradiance incident on a cell each month in kWh/m². It can be seen that months where tandem cells exhibit the highest energy yields (May and July) do not correspond with months where the greatest total energy was incident on the cell (March and June) i.e. the energy yield per month is not directly proportional to the total irradiance incident on the cell. This may seem surprising since the efficiency of solar cells is typically directly proportional to the intensity of light incident on it. However, this only applies if the spectrum of light contributing to the overall intensity can be absorbed by the cell. There is a further complication for tandem cells in that the efficiency of the device will suffer unless the spectrum contributing to the increased intensity is balanced in such a way as to allow current matching between layers. Figure 3 therefore suggests that spectral variations throughout the year have a significant effect on the tandem cells analysed here and this will be discussed in more detail in the next section.

### 3.3. *Spectral effects*

The difference between the predicted annual energy yield and AM1.5G efficiency for each tandem cell considered is between 2.4 - 2.8% (absolute) over the top cell bandgap range (Table II). The fact there is not a greater difference between the yields of the tandem cells with narrower or wider bandgap top cells suggests that the deviations in spectrum from the AM1.5G standard over a calendar year affect all cells in a similar manner. To analyse this further we show the average spectrum for the months with the highest and lowest monthly energy yield (May and December respectively) in Figure 4, normalised at 700 nm, and compare them to the AM1.5G spectrum. It is clear there are significant differences between these average spectra with a much stronger short wavelength (< 650 nm) intensity obvious in May and slightly larger long wavelength intensity in December. Described by their average photon energy (APE), the May and December spectra have APE values of 1.88 and 1.83 respectively in the 350-1050 nm range, where the AM1.5G spectrum has an APE of 1.88 over the same range [23]. Also shown for comparison are the bandgaps considered for the top cells, that are all greater than 650 nm. This means that all the III-V top cells analysed absorb the wavelengths of light where the maximum variation in spectrum occurs during the year, i.e. the variations in spectrum affect all top cells, regardless of their bandgap, leading to similar deviations in actual energy yield compared to AM1.5G efficiency.

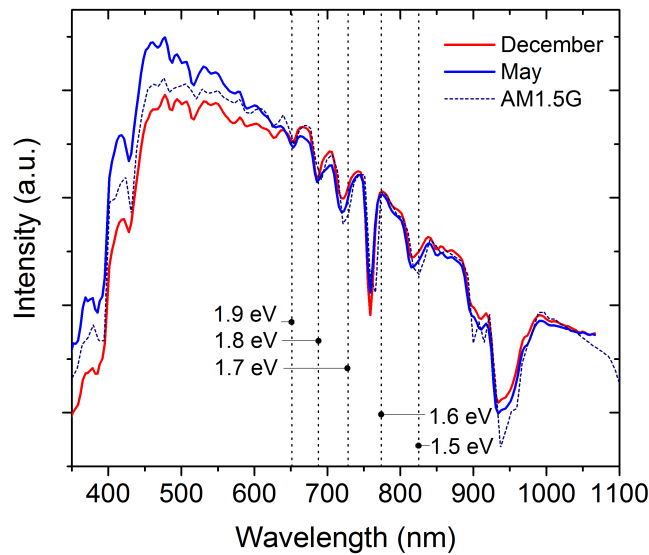

Figure 4: Comparison of the average spectrum for May and December alongside AM1.5G.

Spectral effects play a more interesting role in the difference in energy yield between months. Figure 4 shows that cell designs optimised using the AM1.5G reference spectrum, will be on average current-limited by the bottom cell in May. This is because the increased short wavelength activity in May (compared to the AM1.5G spectrum) causes an increase in current generation in the top cell that is not matched by a similar increase in the silicon layer. In December the top cell will be current limiting due to the decreased intensity of the shorter wavelengths and slight increase of longer wavelength intensity compared to the AM1.5G standard. To analyse these effects further, we calculated the APE of every spectrum used throughout the year as shown in Fig. 5(a) and in Figure 5(b) we show the ratios between the top and bottom cell currents ($J_{III-V}/J_{Si}$) at each time step throughout the year for the 1.8 eV top cell case. The current ratios can be seen to closely follow the APE trend. APE values of < 1.88 eV coincide with current ratios <1 i.e. for red-shifted spectra the cell is top-cell limited, while for APE values of > 1.88eV the current ratio is typically >1. During March to the end of August (days 91 – 243), it is the silicon cell that predominately limits the power produced by the complete tandem cell, with the effect most significant impact during morning and evening hours when the 1.8 eV top cell current can be up to 1.5 times that of the silicon cell. For the rest of the year, principally during morning and evening hours, the 1.8 eV top cell is current limiting.

Figure 5(b) essentially represents the extent to which the ratio of photon absorption in the III-V and silicon layers match what is achieved under the AM1.5G standard spectrum. Deviations in this ratio will cause non-optimal performance of the tandem cell, leading to current mismatch and, in Figure 5(b), a current ratio not equal to 1. For a large proportion of the year, the current ratio varies between 0.9 and 1.1, while the APE remains between 1.82-1.92, indicating that the ratio of photon absorption in each layer of the tandem cell remained close to the AM1.5G conditions during those periods. This observation is reflected in the limited 2.4 to 2.8% (absolute) decrease in annual energy yield compared to AM1.5G efficiency shown in Table II. This implies that for the majority of daylight hours each tandem cell works close to its capacity and this is confirmed in Figure 5(c) where efficiencies of >25% are shown for a large proportion of the year.

Figure 5(c) shows that for May (Days 121-151), the spectral changes in the morning and evening are least pronounced, with a high efficiency of >25% achieved throughout the day. In June (days 152-181), however, the morning blue-shift is so strong that for the first two hours of most days the silicon cell is strongly current limiting. This reduces the energy produced over a full day compared to May, despite longer daylight hours, and this explains the lower monthly % energy yield for June (shown in Figure 3). The low monthly % yield in December can be explained by the simple fact there are the least daylight hours so the overall yield is more significantly impacted by low efficiency in the evening and morning. There is clearly a much shorter period where the simulated tandem cell efficiency is >25%, i.e. on average high efficiency is only achieved between 9:00 am and 15:00 pm. What the analysis shows is that for future PV systems using tandem cell modules, cell designs will be possible that optimise the energy production for certain periods during the year i.e. increasing winter or summer response. Detailed analysis to determine the configuration of tandem cells that are optimized for different times of the year would be a very interesting avenue for future work that could be investigated with the model presented here. Fully exploring this parameter space would involve a significant amount of modelling effort but the results would be of interest to the solar cell community.

Up to this point our study has focused on single-pass absorption in the bottom Si cell, as we consider a flat clean surface to be preferential for most III-V/Si integration strategies such as wafer-bonding and metamorphic growth. In an attempt to further this study, light-trapping in the Si bottom cell was investigated. Light-trapping on the front surface of the Si cell in a III-V/Si tandem cell is possible and has previously been shown experimentally [9]. Our calculations show an increase of ~3% in AM1.5G efficiency is possible in tandem solar cell efficiency considering Lambertian light-trapping in the bottom cell. This significant increase is due to long-wavelength absorption in Si and this effect can be effectively captured by the model when the input spectral data covers a wide range of wavelengths. This is the case with the AM1.5G spectrum and so the calculated efficiency gains with light-trapping is expected. The extended model was applied to the annual energy yield simulations but the results were not conclusive due to the limitations in the data source used, where data for wavelengths >1066 nm is not given. While the error this causes for single-pass absorption is low and tolerable, it reduces the accuracy of energy yield simulations considering light-trapping since the spectral content of the measured input data does not include enough information in the long-wavelength range for benefits of light-trapping to be fully investigated. For a comprehensive study of the effect of the electrical and optical properties of the bottom Si device on tandem solar cell efficiency under the AM1.5G spectrum, see Green et al. [13]. Our future analysis will consider light-trapping in the Si bottom cell to look at the effect of higher currents from the bottom device on tandem cell energy yields.

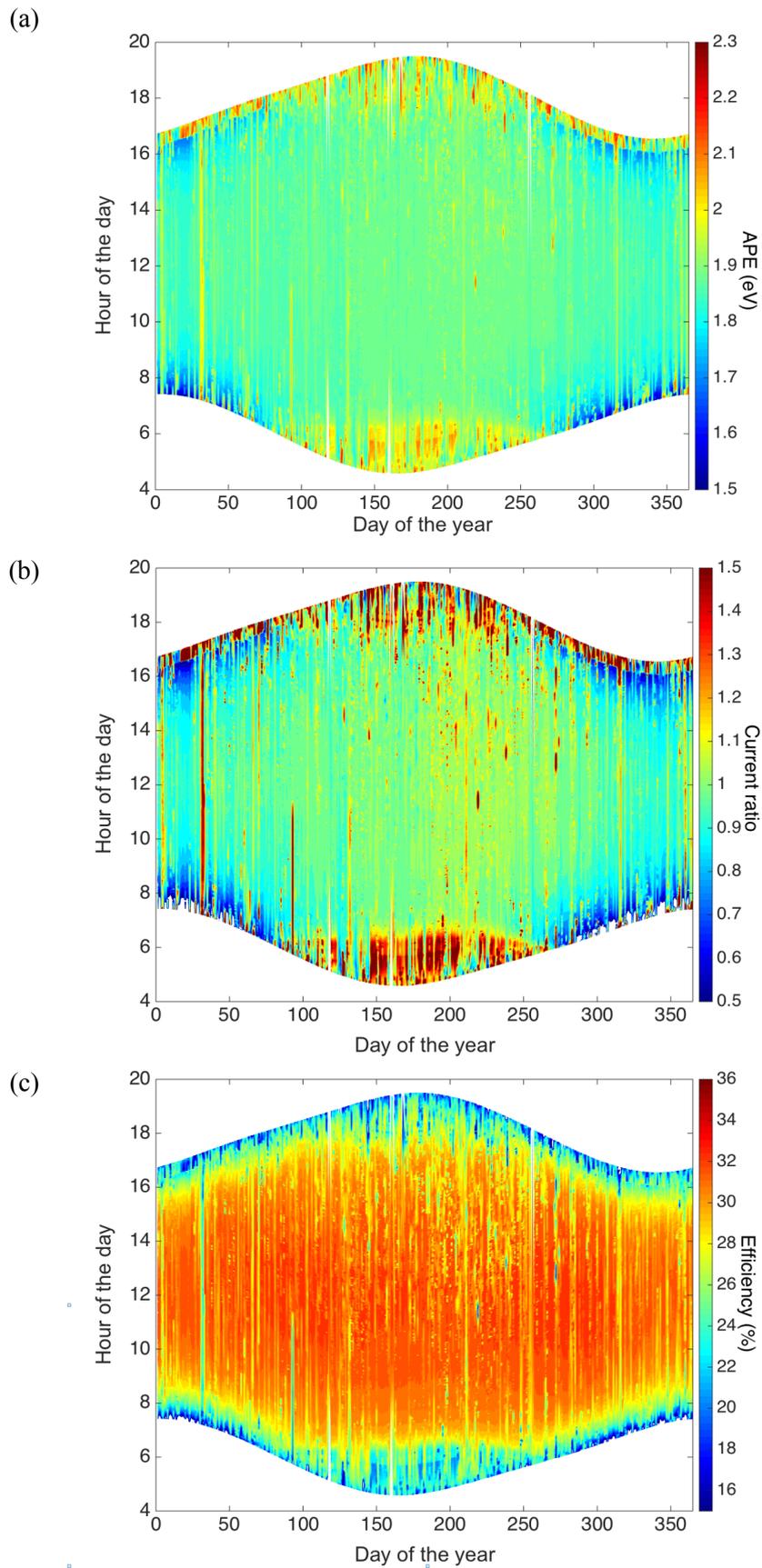

Figure 5: (a) The APE of the solar spetra used in this analysis for the full calendar year (b) Ratio of the photocurrents produced by a 1.8 eV top cell and silicon bottom cell over the year (c) Simulated efficiency of the 2-terminal tandem solar cell with a 1.8 eV bandgap top cell for each timestep considered over the year.

*3.4. Comparison to 4-terminal approach*

Providing separate contacts to each cell in a tandem stack (Figure 1(b)) is one approach to reduce spectral sensitivity and increase annual energy yield in tandem cells [24] as it enables different currents to be produced per sub-cell without one limiting the other. Here we compare the average monthly energy yield of a tandem cell with a 1.6 eV top cell (similar to the bandgap of $CH_3NH_3PbI_3$ perovskite solar cells [9]) series-connected to a silicon bottom cell to the same tandem cell with parallel 4-terminal connections (Figure 6). A similar tandem cell with a 1.8 eV top cell was also analysed. In the 4-terminal case the top cell thickness can be maximised as current-matching is no longer required between cells. The thickness is fixed at 3 μm, which is a realistic maximum thickness likely to be manufactured.

Figure 6 shows that providing additional terminals in the 1.8 eV case only marginally improves the performance over the year, with predicted annual energy yields of 542 and 548 kWh/m$^2$ for the 2- and 4-terminal cases respectively. It is clear that the optimised thickness of the 1.8 eV top cell in the 2-terminal case results in high efficiency throughout the year. Comparing Figure 5 and Figure 7 shows that the 4-terminal device mainly improves the efficiency during morning and evening hours and so reduces the effect of the blue or red-shifted spectra during these hours. However, the low intensity during these times results in little change in the overall annual energy yield and leads to a marginal increase of only 6 kWh/m$^2$ over the full year. The situation is very different for the non-optimal 1.6 eV case where a significant difference is found between the 2- and 4-terminal cases where the predicted annual energy is 487 and 539 kWh/m$^2$ respectively, similar to the difference found for 2-terminal and 4-terminal GaAs-Si tandem cells in [14]. While stacking solar cells in 4-terminal configuration may be more straightforward from a manufacturing perspective, this result shows that with an optimum bandgap top cell, a 2-terminal approach yields approximately the same energy as a 4-terminal device over the course of the year and so may be more appealing from a cost perspective.

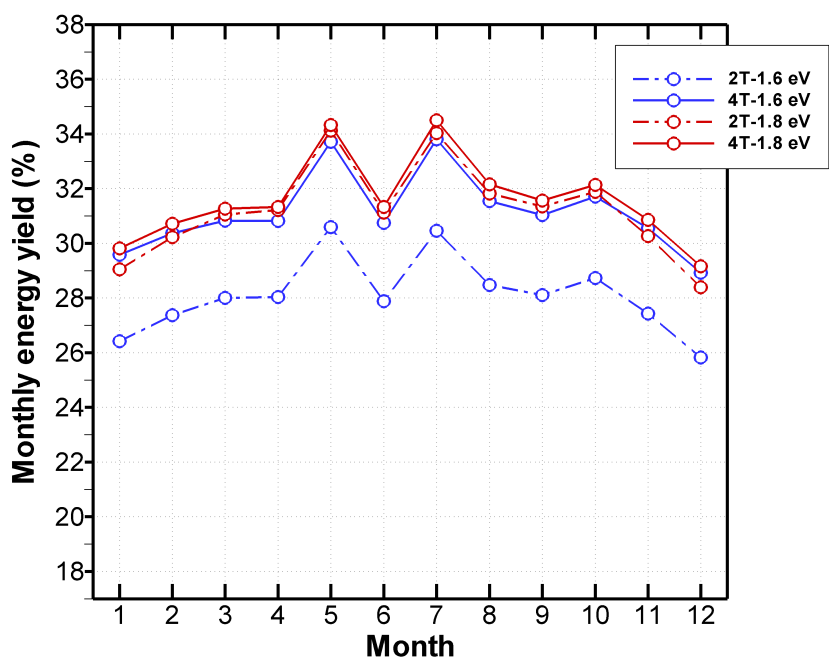

Figure 6: The predicted monthly energy yield (%) of 2-terminal and 4-terminal tandem solar cells with 1.6 and 1.8 eV top cells.

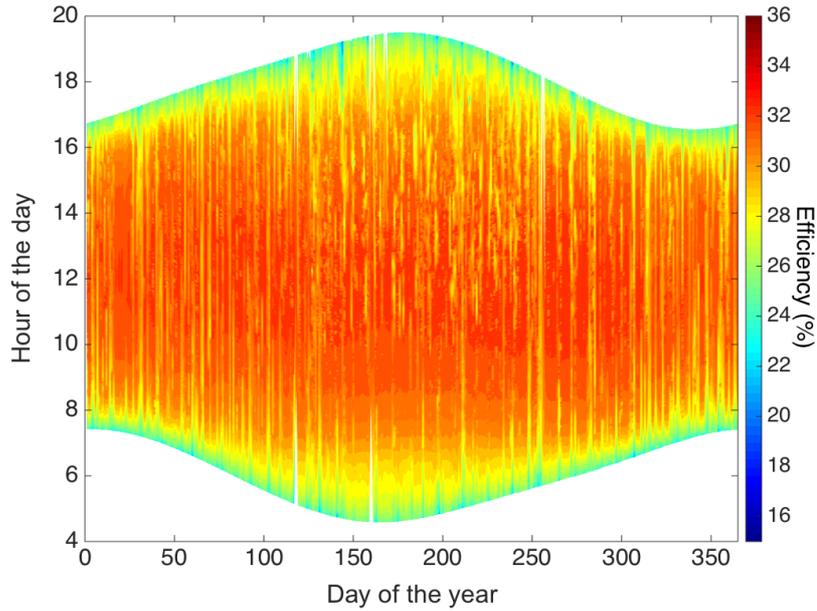

Figure 7: Simulated efficiency at each time step over the year for a 4-terminal tandem cell comprising a 1.8 eV top cell bandgap, which can be compared to the 2-terminal device analysed in Figure 5(b).

## 4. Conclusions

The need to increase the efficiency of current silicon solar cells will likely predicate the use of tandem cells involving a wide-bandgap top cell made from III-V or Perovskite semiconductors. Typical tandem cell designs involve precise current-matching between the sub-cells to ensure optimum photon absorption and to maximise the current produced in a series-connected cell. This optimisation is often carried out considering static standard spectra, such as AM1.5G. This begs the question of how such optimised designs will perform under realistic spectra experienced in actual deployments, where the spectrum will likely deviate from the optimum AM1.5G conditions and continuously degrade the efficiency of the tandem cell. We have used measured spectral data from Denver, Colorado, to show the annual energy yield of tandem cells can be less than the predicted efficiency derived using AM1.5G, however, the average difference was in the order of only 2.8% (absolute), despite the continuous deviations from optimum spectral conditions over the year. It was found that the difference between the annual yield and predicted efficiency remains similar irrespective of the bandgap of the top cell. This is attributed to any differences in the spectrum throughout a day, month and year typically being at wavelengths absorbed by the top cell. Current mismatch is most significant during morning and evening hours, which leads to considerably reduced efficiency at these times. However, this does not have a considerable effect on the annual energy yield (absolute) as the morning and evening spectra are low power and there is minimal energy to convert. Further analysis is required to determine if these results will hold true for other global locations and for spectral data from different years.

Devices with 4-terminals were also analysed as this configuration avoids suboptimal current-mismatch in the different cells. A significant observation from our analysis is the minimal difference between the annual energy yield of a tandem cell consisting of a 1.8 eV optimised top cell whether a 2-terminal series-connected or 4-terminal parallel connection is used. This indicates that tandem cells should be designed with the simplest and most cost effective fabrication method, as the terminal connection strategy is not critical for providing higher yields. This analysis does not hold true for non-optimal top cell bandgaps of 1.6 eV, which result in significantly different yields from 4- and 2-terminal designs.

## Acknowledgements

The authors wish to acknowledge the financial assistance of the Irish Industrial Development Agency.

## References

[1] M.A. Green, K. Emery, Y. Hishikawa, W. Warta, E.D. Dunlop, Solar cell efficiency tables (version 46), Prog. Photovolt. Res. Appl. 23 (2015) 805–812. doi:10.1002/pip.2637.
[2] ITRPV Eighth Edition 2017, ITRPV, n.d.
[3] K. Masuko, M. Shigematsu, T. Hashiguchi, D. Fujishima, M. Kai, N. Yoshimura, T. Yamaguchi, Y. Ichihashi, T. Mishima, N. Matsubara, T. Yamanishi, T. Takahama, M. Taguchi, E. Maruyama, S. Okamoto, Achievement of More Than 25% Conversion Efficiency With Crystalline Silicon Heterojunction Solar Cell, IEEE J. Photovolt. 4 (2014) 1433–1435. doi:10.1109/JPHOTOV.2014.2352151.
[4] D.D. Smith, P. Cousins, S. Westerberg, R.D. Jesus-Tabajonda, G. Aniero, Y. Shen, Toward the Practical Limits of Silicon Solar Cells, IEEE J. Photovolt. Early Access Online (2014). doi:10.1109/JPHOTOV.2014.2350695.
[5] International Technology Roadmap for Photovoltaic - 2014 Results, 2015. http://www.itrpv.net/Reports/Downloads/2015/ (accessed August 24, 2015).
[6] F. Dimroth, M. Grave, P. Beutel, U. Fiedeler, C. Karcher, T.N.D. Tibbits, E. Oliva, G. Siefer, M. Schachtner, A. Wekkeli, A.W. Bett, R. Krause, M. Piccin, N. Blanc, C. Drazek, E. Guiot, B. Ghyselen, T. Salvetat, A. Tauzin, T. Signamarcheix, A. Dobrich, T. Hannappel, K. Schwarzburg, Wafer bonded four-junction GaInP/GaAs//GaInAsP/GaInAs concentrator solar cells with 44.7% efficiency, Prog. Photovolt. Res. Appl. 22 (2014) 277–282. doi:10.1002/pip.2475.
[7] I. Mathews, D. O'Mahony, K. Thomas, E. Pelucchi, B. Corbett, A.P. Morrison, Adhesive bonding for mechanically stacked solar cells, Prog. Photovolt. Res. Appl. 23 (2015) 1080–1090. doi:10.1002/pip.2517.
[8] T. Matsui, H. Sai, T. Suezaki, M. Matsumoto, K. Saito, I. Yoshida, M. Kondo, Development of highly stable and efficient amorphous silicon based solar cells, in: Proc 28th Eur. Photovolt. Sol. Energy Conf. 2013, Paris, France, 2013: pp. 2213–2217.
[9] J.P. Mailoa, C.D. Bailie, E.C. Johlin, E.T. Hoke, A.J. Akey, W.H. Nguyen, M.D. McGehee, T. Buonassisi, A 2-terminal perovskite/silicon multijunction solar cell enabled by a silicon tunnel junction, Appl. Phys. Lett. 106 (2015) 121105. doi:10.1063/1.4914179.
[10] S. Essig, M.A. Steiner, C. Allebé, J.F. Geisz, B. Paviet-Salomon, S. Ward, A. Descoeudres, V. LaSalvia, L. Barraud, N. Badel, A. Faes, J. Levrat, M. Despeisse, C. Ballif, P. Stradins, D.L. Young, Realization of GaInP/Si Dual-Junction Solar Cells With 29.8% 1-Sun Efficiency, IEEE J. Photovolt. 6 (2016) 1012–1019. doi:10.1109/JPHOTOV.2016.2549746.
[11] J.V. Holm, M. Aagesen, Y. Zhang, J. Wu, S. Hatch, H. Liu, Bandgap optimized III-V (GaAsP) nanowire on silicon tandem solar cell, device and data, in: Photovolt. Spec. Conf. PVSC 2014 IEEE 40th, 2014: pp. 1041–1044. doi:10.1109/PVSC.2014.6925092.
[12] T.J. Grassman, J.A. Carlin, C. Ratcliff, D.J. Chmielewski, S.A. Ringel, Epitaxially-grown metamorphic GaAsP/Si dual-junction solar cells, in: Photovolt. Spec. Conf. PVSC 2013 IEEE 39th, 2013: pp. 0149–0153. doi:10.1109/PVSC.2013.6744117.
[13] I. Almansouri, A. Ho-Baillie, S.P. Bremner, M.A. Green, Supercharging Silicon Solar Cell Performance by Means of Multijunction Concept, IEEE J. Photovolt. 5 (2015) 968–976. doi:10.1109/JPHOTOV.2015.2395140.
[14] H. Liu, Z. Ren, Z. Liu, A.G. Aberle, T. Buonassisi, I.M. Peters, The realistic energy yield potential of GaAs-on-Si tandem solar cells: a theoretical case study, Opt. Express. 23 (2015) A382. doi:10.1364/OE.23.00A382.
[15] S.P. Philipps, G. Peharz, R. Hoheisel, T. Hornung, N.M. Al-Abbadi, F. Dimroth, A.W. Bett, Energy harvesting efficiency of III–V triple-junction concentrator solar cells under realistic spectral conditions, Sol. Energy Mater. Sol. Cells. 94 (2010) 869–877. doi:10.1016/j.solmat.2010.01.010.
[16] S.R. Kurtz, J.M. Olson, D.J. Friedman, J.F. Geisz, A.E. Kibbler, K.A. Bertness, Passivation of Interfaces in High-Efficiency Photovoltaic Devices, in: Proc. MRS Spring Meet., San Francisco, California, 1999.
[17] M.A. Green, M.J. Keevers, Optical properties of intrinsic silicon at 300 K, Prog. Photovolt. Res. Appl. 3 (1995) 189–192. doi:10.1002/pip.4670030303.

[18] N.L.A. Chan, N.J. Ekins-Daukes, J.G.J. Adams, M.P. Lumb, M. Gonzalez, P.P. Jenkins, I. Vurgaftman, J.R. Meyer, R.J. Walters, Optimal Bandgap Combinations—Does Material Quality Matter?, IEEE J. Photovolt. 2 (2012) 202–208. doi:10.1109/JPHOTOV.2011.2180513.
[19] K.-H. Lee, K. Araki, L. Wang, N. Kojima, Y. Ohshita, M. Yamaguchi, Assessing material qualities and efficiency limits of III–V on silicon solar cells using external radiative efficiency, Prog. Photovolt. Res. Appl. 24 (2016) 1310–1318. doi:10.1002/pip.2787.
[20] J. Zhao, A. Wang, M.A. Green, F. Ferrazza, 19.8% efficient "honeycomb" textured multicrystalline and 24.4% monocrystalline silicon solar cells, Appl. Phys. Lett. 73 (1998).
[21] T.P. White, N.N. Lal, K.R. Catchpole, Tandem Solar Cells Based on High-Efficiency c-Si Bottom Cells: Top Cell Requirements for #x003E;30 #x0025; Efficiency, IEEE J. Photovolt. 4 (2014) 208–214. doi:10.1109/JPHOTOV.2013.2283342.
[22] NREL, NREL Solar Radiation Research Laboratory Baseline Measurement System, Https://Www.nrel.gov/Midc/Srrl_bms/. (n.d.). https://www.nrel.gov/midc/srrl_bms/.
[23] M. Norton, A.M.G. Amillo, R. Galleano, Comparison of solar spectral irradiance measurements using the average photon energy parameter, Sol. Energy. 120 (2015) 337–344. doi:10.1016/j.solener.2015.06.023.
[24] Y. Mols, L. Zhao, G. Flamand, M. Meuris, J. Poortmans, Annual energy yield: A comparison between various monolithic and mechanically stacked multijunction solar cells, in: 2012 38th IEEE Photovolt. Spec. Conf. PVSC, 2012: pp. 002092–002095. doi:10.1109/PVSC.2012.6318010.